\documentclass[aps,prl,twocolumn,footinbib,superscriptaddress,amsmath,amssymb]{revtex4-1} 

\usepackage{graphicx}
\usepackage{subfigure}
\usepackage{epsfig} 
\usepackage{dcolumn}
\usepackage{bm}
\usepackage{times}
\usepackage[normalem]{ulem}
\usepackage{color}

\begin{document}

\title{Anomalous near-field heat transfer in carbon-based nano-structures with edge states}

\author{Gaomin Tang}
\email{gmtang1212@gmail.com}
\affiliation{Department of Physics, National University of Singapore, Singapore 117551, Republic of Singapore}
\author{Han Hoe Yap}
\email{e0095832@u.nus.edu}
\affiliation{NUS Graduate School for Integrative Sciences and Engineering, Singapore 117456, Republic of Singapore}
\affiliation{Department of Physics, National University of Singapore, Singapore 117551, Republic of Singapore}
\author{Jie Ren}
\email{xonics@tongji.edu.cn}
\affiliation{Center for Phononics and Thermal Energy Science, China-EU Joint Center for Nanophononics, Shanghai Key Laboratory of Special Artificial Microstructure Materials and Technology, School of Physics Science and Engineering, Tongji University, 200092 Shanghai, China}
\author{Jian-Sheng Wang}
\email{phywjs@nus.edu.sg}
\affiliation{Department of Physics, National University of Singapore, Singapore 117551, Republic of Singapore}

\date{\today}

\begin{abstract}
We find an unusually optimal near field heat transfer, where the maximum heat transfer is reached at experimentally feasible gap separation. We attribute this to the localized zero-energy electronic edge states, which also substantially changes the near-field behaviors. We demonstrate these anomalous behaviors in two typical carbon-based nano-structures: zigzag single-walled carbon nanotubes and graphene nano-triangles. For the system of carbon nanotubes, the maximal heat flux in this work surpasses all the previous results reported so far by several orders of magnitude. The underlying mechanisms for the peculiar effects are uncovered from a simple Su-Schrieffer-Heeger model. Our findings also offer a novel route to active near-field thermal switch, where the heat flux can be modulated through tuning the presence or absence of edge states.
\end{abstract}

\maketitle

{\it Introduction.--}
Heat transfer in the far field can be well described by Planck's theory of black-body radiation \cite{Planck} and obeys the Stefan-Boltzmann law, which is independent of gap separation distance. 
When the gap separation between two bodies becomes smaller than Wien's wavelength, heat transfer in the near field becomes distance dependent and has been demonstrated to be much larger than that in the far field \cite{near1, review3, review4}. 
Within the fluctuational electrodynamics \cite{review3, review4, PvH}, the near-field heat flux typically increases as the two bodies become closer. As such, great efforts have been dedicated to reducing the gap sizes from orders of $1\,\mu$m \cite{XJB, mu-m1, mu-m2} to a few nanometers \cite{nm2, nm3, nm4, nm5-PvH} or even down to few \textup{\AA}ngstr\"{o}ms \cite{angstrom1, angstrom2}, resulting in several folds to several orders of heat transfer enhancement compared to the corresponding far-field results, which may prove useful for near-field thermal management. To our best knowledge, heat flux typically increases with the decrease of gap separation.  

	Besides by reducing gap separation, several other approaches have been brought forward to enhance near-field heat transfer. Pendry showed that the heat flux can be greatly enhanced by tuning the resistivity of the material \cite{resist1}. Covering both surfaces with adsorbates, so that resonant photon tunneling happens between adsorbate vibrational modes, can also enhance the heat flux \cite{adsorbate}. 
Two types of surface waves, which propagate along the material-vacuum interfaces, have been mainly used to increase heat transfer in the near field. One is surface phonon polariton supported in polar dielectrics, such as SiC and SiO$_2$ \cite{SPhP1, SPhP3, SPhP-SPP1, SPhP-SPP2}. The other type is surface plasmon polariton on materials supporting low frequency  plasmon \cite{adsorbate, SPhP-SPP1, SPhP-SPP2, plasmon1,plasmon2}, such as graphene \cite{graphene1, graphene2, Yu}, black phosphorus \cite{bp} and silicon \cite{silicon1, silicon2}. 
Using hyperbolic metamaterials can help to enhance heat transfer as well \cite{hyperbolic1,hyperbolic2,hyperbolic4}.



\begin{figure}[tb!]
\centering
\includegraphics[width=\columnwidth]{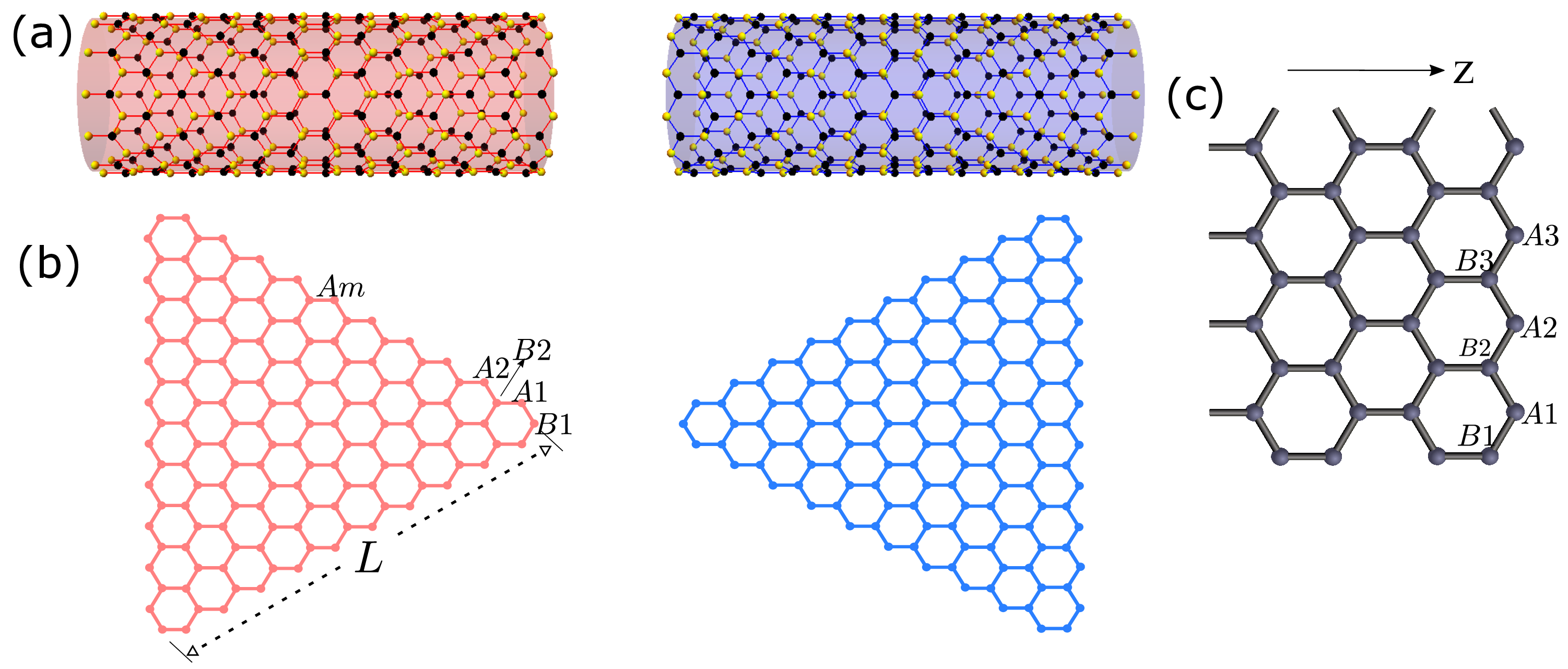} \\ 
\caption{Near-field heat transfer between two SWCNTs (a) and graphene nano-triangles forming a bowtie shape (b) which are separated by a vacuum gap with distance $d$. (c) Lattice structure of a zigzag SWCNT corresponding to the left side in panel (a). For the graphene nano-triangles, which are equilateral, we show the case with side length to be $L=4.26\,$nm. }
\label{fig1}
\end{figure}

	In this letter, we report peculiar vacuum gap dependence and enhancement of heat transfer in the near field in the presence of electronic edge states. We consider heat transfer between two carbon-based nano-structures harboring electronic edge states separated by a vacuum gap no further than $20\,$nm, so that (i) the electron-electron interaction dominates the heat transfer, and (ii) an atomistic description is more appropriate. The heat current expression is given in the formalism of the nonequilibrium Green's function (NEGF) within the random phase approximation. Heat current can reach maximum at a finite vacuum gap if the real part of the charge susceptibility is large near zero angular frequency, and this can be realized through the presence of edge states. 
	The peculiar behaviors are demonstrated using zigzag single-walled carbon nanotubes (SWCNTs) [See Fig.~\ref{fig1}(a)] and graphene nano-triangles forming a bowtie shape [Fig.~\ref{fig1}(b)]. We uncover the the mechanism using the simple Su-Schrieffer-Heeger (SSH) chains [See the inset of Fig.~\ref{fig3}(b)].

{\it Theoretical formalism.--}
When two metallic surfaces are separated by a vacuum gap, which is much smaller than the Wien's wavelength $\lambda_{th}$ (several micrometers at room temperature), the contribution to heat transfer from the retarded vector potential can be safely ignored, and the electron-electron interaction dominates the heat transfer \cite{Yu,Mahan,JS1,JS3,JS4,GM1, supp}. If the vacuum gap is below around $1.5\,$nm, electron tunneling process can happen \cite{JS3, crossover}, the picture dominated by electron-electron interaction does not apply. So our formalism below can faithfully describe the heat transfer with the vacuum gap $d$ in the range $1.5\,{\rm nm} < d \ll \lambda_{th}$. The lattice Hamiltonian of a general heat transfer mediated by electron-electron interaction can be written as, 
\begin{equation}
H = \sum_{mn} c_m^\dag h_{mn} c_n+ \frac{e_0^2}{2}\sum_{mn} c_m^\dag c_m v_{mn} c_n^\dag c_n , \label{Hamiltonian}
\end{equation}
with $e_0$ the elementary charge. $c_m$ ($c_m^\dag$) is the fermionic annihilation (creation) operator of lattice site $m$ on the left or right side, and $h_{mn}$ is the on-site energy for $m=n$ and hopping parameter for $m\neq n$ locating on the same side. For the situation of indices $m$ and $n$ locating on different sides, $h_{mn}$ vanishes. This implies that electron tunneling from one side to the other is impossible when the vacuum gap is far greater than the spacing between nearest-neighbor atoms. $v_{mn}$ is the Coulomb potential between site $m$ and $n$. Heat transfer occurs via the charge fluctuations between the electronic states sitting at the edges from both sides for the setups considered.

Under random phase approximation, the heat current is expressed as \cite{Yu,JS4},
\begin{equation}
J = \int_0^{\infty} \frac{d\omega}{2\pi} \hbar\omega {\cal T}(\omega)\big[N_L(\omega)-N_R(\omega)\big] ,
\end{equation}
where $N_\alpha(\omega) = 1/[e^{\beta_\alpha \hbar\omega}-1]$ is the Bose-Einstein distribution with $\beta_\alpha = 1/(k_B T_\alpha)$ and $\alpha=L,R$. 
The spectral transfer function is given by, 
\begin{equation}
{\cal T}(\omega) = 4{\rm Tr} \big\{ \Delta^\dag(\omega) v_{RL} {\rm Im}[\chi_L(\omega)] v_{LR} \Delta(\omega) {\rm Im}[\chi_R(\omega)] \big\} ,
\end{equation}
where the trace is over lattice sites. $\Delta(\omega)$ is expressed as
\begin{equation}
\Delta(\omega) = [{\bf I}- \chi_R(\omega) v_{RL} \chi_L(\omega) v_{LR}]^{-1} , \label{Delta}
\end{equation}
with identity matrix ${\bf I}$ and the charge susceptibility $\chi_{\alpha}(\omega)$ in lattice space obtained through electronic Green's function \cite{supp}. 
The entries of the Coulomb potential matrices between left and right sides are $v_{mn} = 1/(4\pi\epsilon_0 d_{mn})$, where $\epsilon_0$ is the dielectric constant of vacuum. $d_{mn}$ is the Euclidean distance between site $m$ and $n$, which sit on different surfaces. The formalism based on NEGF here has been shown \cite{Yu,JS3,JS4} to reduce to that by the fluctuational electrodynamics \cite{PvH,review3, review4} in the non-retardation limit and be equivalent to that given by Mahan \cite{Mahan}.

\begin{figure}[tb!]
\centering
\includegraphics[width=\columnwidth]{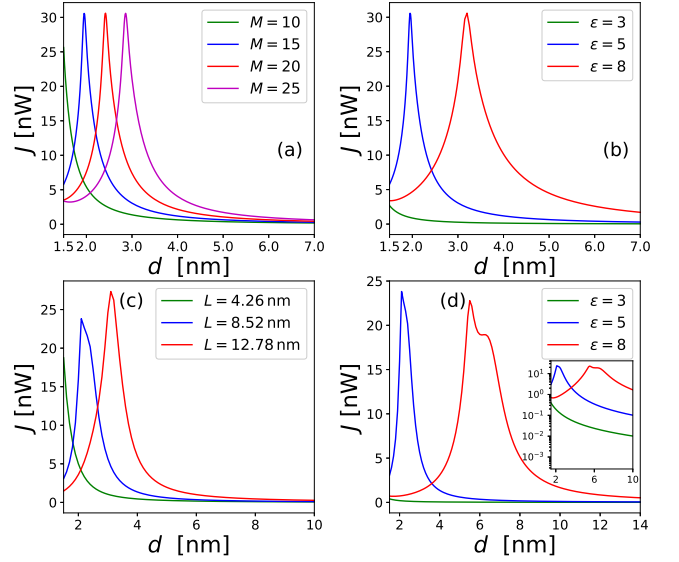}\\
\caption{
Gap separation dependences of heat current between zigzag SWCNTs for different $M$ with $\varepsilon =5$ and $\eta=25\,$meV [panel~(a)], and for different dielectric constants $\varepsilon$ with $M=15$ and $\eta=25\,$meV [panel~(b)]. Gap separation dependence of heat current between graphene nano-triangles for different $L$ with the dielectric constant $\varepsilon =5$ [panel (a)], and for $\varepsilon$ with $L=8.52\,$nm [panel (b)].} 
\label{fig2}
\end{figure}

{\it Near-field heat transfer between zigzag SWCNTs.--}
For the numerical results presented in this work, temperatures of the left and right sides are set as $T_L=400\,$K and $T_R=300\,$K, respectively. 
We first discuss heat transfer between two zigzag SWCNTs in near field. It has been reported that zero-energy localized states (Fujita's edge states) emerge at the edges of zigzag SWCNTs \cite{CNT1, CNT2, CNT3}.  
As shown in Fig.~\ref{fig1}(c), the outermost carbon atoms are denoted by $Am$ and the other surface atoms by $Bm$, where $m$ is the site index. The contribution of heat transfer between two bodies are mainly from these surface carbon atoms. The local electronic density of states for the atoms $Am$ are sharply peaked at zero-energy as shown in Fig.~1(a) in the Supplementary Material \cite{supp}. 
We use $M$ to denote the total number of the outermost carbon atoms, and it is proportional to the radius of the nanotube. Due to the $O(2)$ rotational invariance of carbon nanotubes, all the carbon atoms of $Am$ and $Bm$ are geometrically equivalent in their respective sides. 
The carbon-carbon bond length is $1.42\,$\textup{\AA}. The recursive Green's function technique \cite{Recursive} is used in getting electronic Green's function of the surface sites, where 
the nearest-neighbour hopping constant is given by $2.5\,$eV. A damping constant $\eta=25\,$meV is included in calculating Green's function to account for possible dissipations by such as electron-phonon interactions. A dielectric constant $\epsilon$ is included in calculating the intra-side Coulomb interaction. 

	In Fig.~\ref{fig2}(a), we plot heat currents versus gap distances for zigzag SWCNTs with different radii. The non-monotonic behavior is found for the cases of $M=15$, $20$, and $25$ shown in Fig.~\ref{fig2}(a). The maximal heat currents are identical for these three cases. We find that the larger the nanotube radii, the longer the distances for achieving corresponding maximal heat current.	The condition(s), under which the heat current can reach maximum with several nanometers of gap separation for zigzag SWCNTs, can be obtained analytically. The detailed argument is provided in the Supplementary Material \cite{supp}. The critical distance is approximately the summation of all entries of the real parts of the charge susceptibility at zero frequency, that is, 
\begin{equation}
d_c \approx -\frac{1}{4\pi\epsilon_0} \sum_{m,n}{\rm Re}[\chi_{mn}(\omega=0)] . \label{d_c}
\end{equation}
Since the terminal sites contribute to the critical distance in an additive way, a small amount of disorder or vacancies will only decrease the critical distance slightly without changing the maximal heat current. 
For the case $M=10$ in Fig.~\ref{fig2}(a), the critical distance appears below $1.5\,$nm as predicted by equation~\eqref{d_c}. For such extremely short distance, there may be contribution to the heat transfer from electron tunneling \cite{JS3, crossover}, an aspect which is not included in our model.


	Gap separation dependence of heat current for different dielectric constants $\varepsilon$ in zigzag SWCNTs is shown in Fig.~\ref{fig2}(b). One can find that the critical distance disappears for the case with $\varepsilon =3$. With increasing dielectric constant $\varepsilon$, the critical distance shifts to larger values. This is because the amplitudes of the real parts of charge susceptibility at zero frequency increase as a consequence of less screening.	
Different dielectric constants can be realized by encapsulating or covering the carbon nanotubes with different dielectric materials. Inserting a dielectric material as a core of the nanotubes can change the dielectric constant as well. 

	The area formed by a zigzag SWCNT with $M=15$ is $1.1\times 10^{-18}\,{\rm m}^2$, so that the corresponding heat flux, i.e., heat current per area, is about $2.8\times 10^{10}\, {\rm W}/{\rm m}^2$ with heat current reaching maximum $J=30.6\,$nW at critical distance. This heat flux is several orders of magnitude larger than those mediated by surface phonon polaritons or surface plasmon polaritons \cite{silicon2}, and almost comparable to that of heat conduction. Since the structure-controlled growth of SWCNTs is rapidly advanced \cite{CNT-exp1, CNT-exp2}, the results here can be are expected to be  experimentally realized in the near future.

{\it Near-field heat transfer between graphene nano-triangles.--}
We further show the non-monotonic behavior of gap separation for near-field heat transfer between two gaphene nano-triangles, which are equilateral. We focus on the vertex-to-vertex geometry forming a bowtie shape as shown in Fig.~\ref{fig1}(b), and length of the nano-triangle's side is $L$. The nearest-neighbour hopping constant is chosen as $2.8\,$eV, and the damping constant $\eta=25\,$meV. The graphene nano-triangles can be maintained at thermal equilibrium through optical pumping or by being attached to additional electrodes. As shown in Fig.~2 in Supplementary Material \cite{supp}, the main contributions to the heat transfer are from sites $Am$ indicated in Fig.~\ref{fig1}(b), which have strongly localized zero-energy states. Gap separation dependent behaviors of heat current by varying side length $L$ and dielectric constant $\varepsilon$ are shown Fig.~\ref{fig2}(c) and (d), respectively. 
Similarly to the behaviors found in Fig.~\ref{fig2}(a) for zigzag SWCNTs, we observe non-monotonic behavior for $L=8.52\,$nm, and $L=12.78\,$nm as shown in Fig.~\ref{fig2}(c). From Fig.~\ref{fig2}(d), we see that increasing the dielectric constant in graphene nano-triangles increases the critical distance as well, which is as shown in Fig.~\ref{fig2}(b).

{\it Near-field heat transfer between SSH chains.--}
We bring forth a theoretical understanding for the peculiar phenomenon observed above using the simple one-dimensional SSH chains, which undergo topological phase transitions by varying the hopping parameters \cite{SSH1}.
The Hamiltonian of a SSH chain for side $\alpha$ is expressed as,
\begin{align}
H_{0\alpha} = & -(1+\lambda_\alpha)t \sum_{n=1}^{N} ( c_{An}^\dag c_{Bn} + {\rm H.c.} )  \notag \\
&-(1-\lambda_\alpha)t \sum_{n=1}^{N-1} ( c_{An+1}^\dag c_{Bn} +{\rm H.c.}) ,
\end{align}
with $N$ the number of lattice sites, and $\lambda_\alpha \in [-1,1]$. We consider the case where heat transfer happens only between two end sites, both of which are labeled as $A1$ as shown in the inset of Fig.~\ref{fig3}(a). 
We set $\lambda_L=\lambda_R=\lambda$ and the hopping constant as $t=2.2\,$eV in the calculation. A damping constant with $\eta = 22\,$meV is added to each site in calculating electronic Green's function \cite{SSH-Green}. 
The energy spectrum of an open SSH chain is shown in the right inset of Fig.~\ref{fig4}(a). 
When $\lambda>0$, SSH chain is in a trivial insulator state without in-gap state, and it is in a metallic state for $\lambda=0$ where the gap closes. However, the gap reopens in the topologically nontrivial region with $\lambda<0$, and zero-energy in-gap states appear when open boundary condition is taken. 

\begin{figure}[tb!]
\centering
\includegraphics[width=\columnwidth]{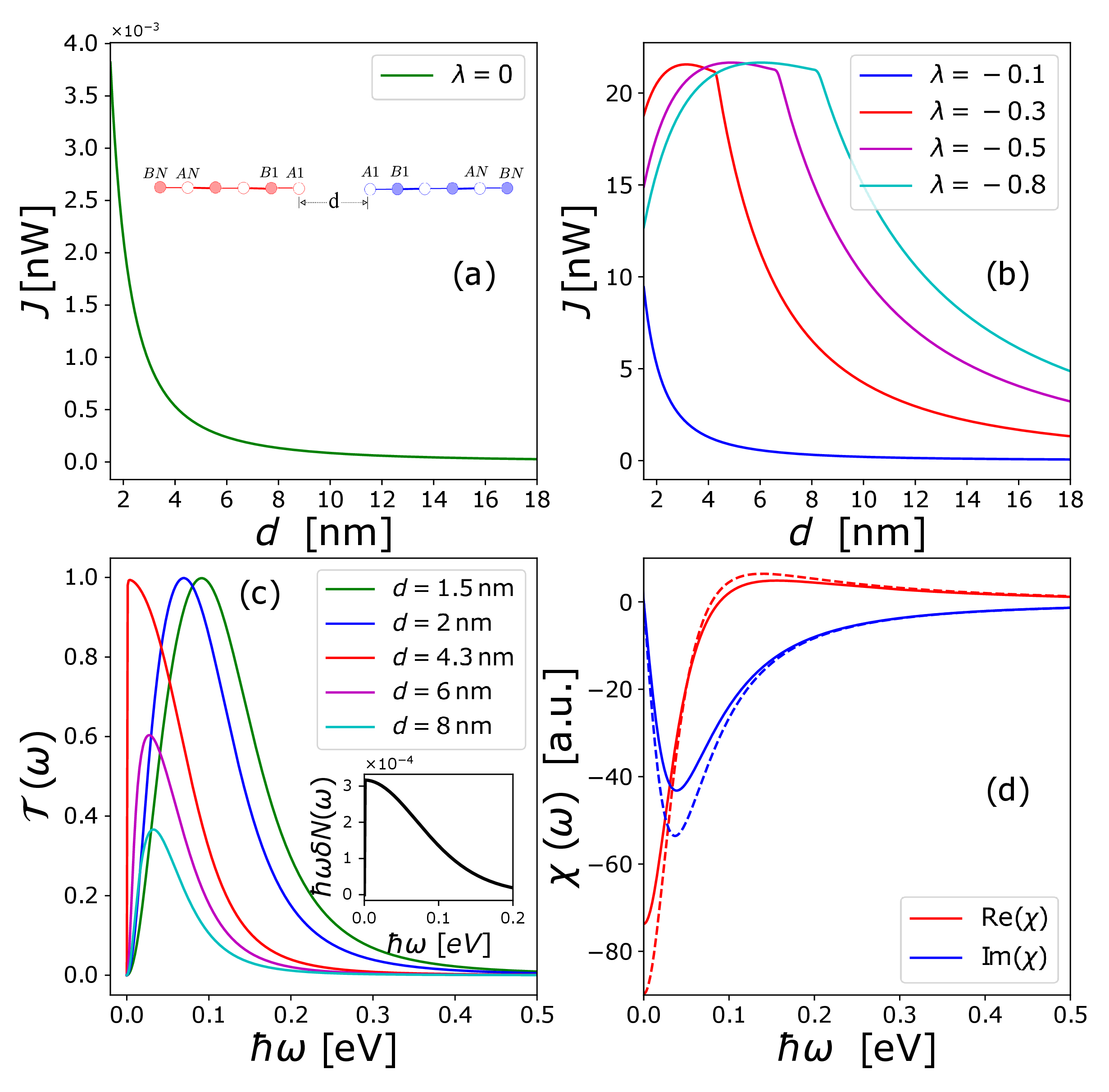} \\ 
\caption{Gap separation dependence of heat current in the metallic phase (a) and in the topologically nontrivial phase with different $\lambda$ (b). Heat currents with gap separation below $1.5\,$nm are not shown because electron tunneling will no longer be negligible and we wish to exclude this effect. (c) The spectral transfer functions for different gap separations $d$ with $\lambda=-0.3$. $d=4.3\,$nm is the critical gap separation above which the current decays with increasing gap separation. (d) Real and imaginary parts of the charge susceptibility function of the end site $A1$ with temperatures $400\,$K ($\chi_L$, solid lines) and $300\,$K ($\chi_R$, dashed lines). 
$\hbar\omega\delta N(\omega)$ is shown as an inset of panel (c). }
\label{fig3}
\end{figure}

The gap separation dependence of heat current in the metallic phase and the topologically nontrivial phase are shown in panels (a) and (b) of Fig.~\ref{fig3}, respectively. For the case of $\lambda=-0.3$, the spectral transfer functions for different gap separations $d$ and the charge susceptibilities $\chi(\omega)$ of the end sites $A1$ are plotted in Fig.~\ref{fig3}(c) and (d), respectively. 
There exists a critical gap distance $d_c$ at which the heat flux achieves its maximum point in the presence of edge state. The explanation of this peculiar distance dependence is as follows. 
In atomic units, the critical distance is approximately equal to the absolute values of the real parts of $\chi_{L/R}(\omega)$ at zero frequency, i.e., $d_c \approx -{\rm Re}[\chi_{L}(\omega=0)]/(4\pi\epsilon_0)$.
(For $\lambda=-0.3$, the critical distance $d_c$ is near $4.3\,{\rm nm}=81.26\,{\rm a.u.}$. The value in atomic unit is between $|{\rm Re}[\chi_L(\omega=0)]|$ and $|{\rm Re}[\chi_R(\omega=0)|$ which are shown in Fig.~\ref{fig3}(d).)
Near the critical distance $d_c$, ${\rm Re}[\chi_L(\omega\rightarrow 0)]v_{LR} \approx {\rm Re}[\chi_R(\omega\rightarrow 0)]v_{RL} \approx -1$. 
One also has ${\rm Im}[\chi_{L/R}(\omega\rightarrow 0)] \approx 0$, so that $v_{LR}\Delta{\rm Im}(\chi_R) |_{\omega\rightarrow 0} \approx i/2$ from equation~\eqref{Delta}, hence we have ${\cal T}(\omega\rightarrow 0)\approx 1$ around the critical distance. Since the function $\hbar\omega \delta N(\omega)$ is a decreasing function with respect to angular frequency $\omega>0$ (shown as an inset of Fig.~\ref{fig3}(c)), the magnitude of the spectral transfer function at low $\omega$ dominates the heat current amplitude. The resonant peak of ${\cal T}(\omega)$ is located close to $\omega=0$ at the critical gap distance at which the heat current achieves its maximum. 
The resonant peak shifts towards larger angular frequency with decreasing gap distance (as shown in Fig.~\ref{fig3}(c) for $\lambda=-0.3$), and this results in a suppressed heat current. Above the critical distance, the condition for the appearance of the resonant peak cannot be satisfied. With increasing distance above the critical distance, the magnitude of the spectral transfer function decreases, so does the heat current. 
In the metallic phase, we have ${\rm Re}[\chi_L(\omega = 0)] \approx -5$, which means that the critical distance is about $2.6\,$\textup{\AA}. At such a small gap distance, heat conduction due to electron tunneling can happen, and our formalism does not apply \cite{JS3}.  
As $\lambda$ approaches $-1$, the critical distance increases due to the fact that edge states become more localized. The maximum heat currents are almost the same in presence of edge states because they share similar spectral transfer function profiles regardless of the critical gap distances. The non-monotonic behaviors with respect to gap separation have been reported for heat radiation between a cylinder and a perforated surface due to dipolar effects \cite{non-monotonic1} and in a multilayer structure due to the interplay of contributions from different surfaces \cite{non-monotonic2}. If the SSH chains experience more dissipation, i.e., larger $\eta$, the edge sates become less localized, and the critical distance becomes shorter [See Fig.~4 in the Supplementary Material \cite{supp}].

\begin{figure}[tb!]
\centering
\includegraphics[width=\columnwidth]{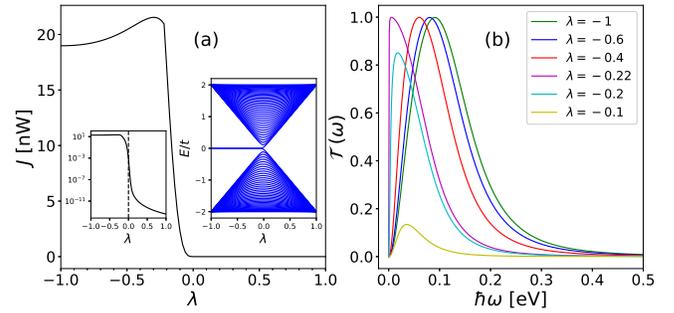} \\ 
\caption{(a) Heat current by changing $\lambda$ for a gap separation of $3\,$nm, at which the electron-electron interactions dominate the heat transfer. A log-scale plot for the heat current is shown in the left inset. (b) The spectral transfer functions for different $\lambda$. Energy spectrum of the SSH chain with $160$ lattice sites as a function of $\lambda$ is shown as a right inset of panel (a). }
\label{fig4}
\end{figure}

In Fig.~\ref{fig4}(a), heat current versus $\lambda$ for gap separation $d=3\,$nm is plotted. The heat current for the metallic phase ($\lambda =0$) is several orders smaller than that for the topologically nontrivial phase. In the trivial insulating phase ($\lambda >0$), heat current is extremely small and almost vanishes. A sharp jump occurs with the phase transition point $\lambda <0$, indicating that the presence of edge state can drastically enhance heat current compared to the metallic phase. The spectral transfer functions for different $\lambda$ are shown in Fig.~\ref{fig4}(b). The increase of heat current as $\lambda$ is changed from $-1.0$ to around $-0.22$ can be attributed to the shift of the resonant peak of the spectral transfer function towards $\omega=0$. This is because that $d=3\,$nm is the critical distance corresponding to $\lambda = -0.22$. At $\lambda = -0.22$, the resonant peak of the spectral function ${\cal T}(\omega)$ is at a frequency close to $\omega=0$. For $\lambda<-0.22$, the chosen gap separation $d=3\,$nm is smaller than the corresponding critical distance, and the resonant peak locates at a larger frequency. 
	By further increasing $\lambda$ from $-0.22$, the SSH chain approaches the metallic phase, and the peak of the spectral transfer function at low angular frequency decreases, thus reducing the heat current. 
The fact that heat current can be greatly enhanced in the presence of edge states can be exploited to design a near-field thermal switch, provided that edge states can be tuned. The discussions of near-field heat transfer between SSH chains, with the same chemical potential applied to both sides, are shown in the Supplementary Material \cite{supp}.

{\it Summary and discussion.--}
We have uncovered the peculiar behaviors of near-field heat transfer in the presence of electronic edge states. Our findings are demonstrated using zigzag SWCNTs and graphene nano-triangles forming a bowtie shape. The underlying mechanism is uncovered through the simple SSH chains. In the presence of localized zero-energy edge states, heat current is greatly enhanced and shows a non-monotonic behavior with respect to vacuum gap separation. The maximal heat flux between zigzag SWCNTs are shown to be extremely large, and surpasses near-field heat flux being reported so far. 

\begin{acknowledgments}
{\it Acknowledgments.--}
The authors thank Jiebin Peng, Giovanni Vignale, Fuming Xu and Songbo Zhang for discussions and comments. G.T. and J.S.W. acknowledge the financial support from RSB funded RF scheme (Grant No. R-144-000-402-114). J.R. is supported by the NNSFC (No. 11775159), Shanghai Science and Technology Committee (No. 18ZR1442800, No. 18JC1410900), and the Opening Project of Shanghai Key Laboratory of Special Artificial Microstructure Materials and Technology.
\end{acknowledgments}


\begin{thebibliography}{99}
\bibitem{Planck} M. Planck, and M. Masius, {\it The theory of heat radiation} (P. Blakiston's Son \& Co, 1914).

\bibitem{near1} C. M. Hargreaves, Phys. Lett. A {\bf 30}, 491-492 (1969).


\bibitem{review3} A. Volokitin and B. Persson, Rev. Mod. Phys. {\bf 79}, 1291 (2007).
\bibitem{review4} B. Song, A. Fiorino, E. Meyhofer, and P. Reddy, AIP Advances {\bf 5}, 053503 (2015). 

\bibitem{PvH} D. Polder and M. Van Hove, Phys. Rev. B {\bf 4}, 3303 (1971).

\bibitem{XJB} J.-B. Xu, K. L\"{a}uger, R. M\"{o}ller, K. Dransfeld, and I. H. Wilson, J. Appl. Phys. {\bf 76}, 7209 (1994). 
\bibitem{mu-m1} R. S. Ottens, V. Quetschke, S. Wise, A. A. Alemi, R. Lundock, G. Mueller, D. H. Reitze, D. B. Tanner, and B. F. Whiting, Phys. Rev. Lett. {\bf 107}, 014301 (2011).
\bibitem{mu-m2} M. Lim, S. S. Lee, and B. J. Lee, Phys. Rev. B {\bf 91}, 195136 (2015).


\bibitem{nm2} A. Kittel, W. M\"{u}ller-Hirsch, J. Parisi, S.-A. Biehs, D. Reddig, and M. Holthaus, Phys. Rev. Lett. {\bf 95}, 224301 (2005).

\bibitem{nm3} L. Worbes, D. Hellmann, and A. Kittel, Phys. Rev. Lett. {\bf 110}, 134302 (2013).

\bibitem{nm4} K. Kloppstech, N. K\"{o}nne, S.-A. Biehs, A. W. Rodriguez, L. Worbes, D. Hellmann, and A. Kittel, Nat. Commun. {\bf 8}, 14475 (2017). 

\bibitem{nm5-PvH} K. Kim, B. Song, V. Fern\'{a}ndez-Hurtado, W. Lee, W. Jeong, L. Cui, D. Thompson, J. Feist, M. T. H. Reid, F. J. Garc\'{i}a-Vidal, J. C. Cuevas, E. Meyhofer, and P. Reddy, Nature, {\bf 528}, 387 (2015). 


\bibitem{angstrom1} V. Chiloyan, J. Garg, K. Esfarjani, and G. Chen, Nat. Commun. {\bf 6}, 6755 (2015).
\bibitem{angstrom2} L. Cui, W. Jeong, V. Fern\'{a}ndez-Hurtado, J. Feist, F. J. Garc\'{i}a-Vidal, J. C. Cuevas, E. Meyhofer, and P. Reddy, Nat. Commun. {\bf 8}, 14479 (2017).

\bibitem{resist1} J. B. Pendry, J. Phys.: Condens. Matter {\bf 11}, 6621 (1999). 

\bibitem{adsorbate} A. I. Volokitin and B. N. J. Persson, Phys. Rev. B {\bf 69}, 045417 (2004).

\bibitem{SPhP1} S. Shen, A. Narayanaswamy, and G. Chen, Nano Lett. {\bf 9}, 2909 (2009).


\bibitem{SPhP3} K. Ito, T. Matsui, and H. Iizuka, Appl. Phys. Lett. {\bf 104}, 10 (2014).

\bibitem{SPhP-SPP1} J. P. Mulet, K. Joulain, R. Carminati, and J. J. Greffet, Microscale Thermophys. Eng. {\bf 6}, 209 (2002).

\bibitem{SPhP-SPP2} H. Iizuka and S. Fan, Phys. Rev. B {\bf 92}, 144307 (2015).

\bibitem{plasmon1} S. V. Boriskina, J. K. Tong, Y. Huang, J. Zhou, V. Chiloyan, and G. Chen, Photonics, {\bf 2}, 659 (2015).

\bibitem{plasmon2} J.-P. Mulet, K. Joulain, R. Carminati, and J.-J. Greffet, Appl. Phys. Lett. {\bf 78}, 2931 (2001).

\bibitem{graphene1} O. Ilic, M. Jablan, J. D. Joannopoulos, I. Celanovic, H. Buljan, and M. Soljacic, Phys. Rev. B {\bf 85}, 155422 (2012).

\bibitem{graphene2} F. V. Ramirez, S. Shen, and A. J. H. McGaughey, Phys. Rev. B {\bf 96}, 165427 (2017).

\bibitem{Yu} R. Yu, A. Manjavacas, and F. J. Garc\'{i}a de Abajo, Nat. Commun. {\bf 8}, 2 (2017). 

\bibitem{bp} Y. Zhang, H.-L. Yi, and H.-P. Tan, ACS Photonics {\bf 5}, 3739 (2018).
2018 5 (9), 3739-3747

\bibitem{silicon1} E. Rousseau, Ma. Laroche, and J.-J. Greffet, Appl. Phys. Lett. {\bf 95}, 231913 (2009).

\bibitem{silicon2} V. Fern\'{a}ndez-Hurtado, F. J. Garc\'{i}a-Vidal, S. Fan, and J. C. Cuevas, Phys. Rev. Lett, {\bf 118}, 203901 (2017).

\bibitem{hyperbolic1} S.-A. Biehs, M. Tschikin, and P. Ben-Abdallah, Phys. Rev. Lett. {\bf 109}, 104301 (2012).

\bibitem{hyperbolic2} S.-A. Biehs, M. Tschikin, R. Messina, and P. Ben-Abdallah, Appl. Phys. Lett. {\bf 102}, 131106 (2013).


\bibitem{hyperbolic4} O. D. Miller, S. G. Johnson, and A. W. Rodriguez, Phys. Rev. Lett. {\bf 112}, 157402 (2014).  

\bibitem{Mahan} G. D. Mahan, Phys. Rev. B {\bf 95}, 115427 (2017).

\bibitem{JS1} J.-S. Wang and J. Peng, Europhys. Lett. {\bf 118}, 24001 (2017).
\bibitem{JS3} Z.-Q. Zhang, J.-T. L\"{u}, and J.-S. Wang, Phys. Rev. B {\bf 97}, 195450 (2018).
\bibitem{JS4} J.-S. Wang, Z.-Q. Zhang, and J.-T. L\"{u}, Phys. Rev. E {\bf 98}, 012118 (2018). 
\bibitem{GM1} G. Tang and J.-S. Wang, Phys. Rev. B {\bf 98}, 125401 (2018).

\bibitem{supp} See Supplemental Material at http://link.aps.org/supplemental/XXXX. 

\bibitem{crossover} R. Messina, S.-A. Biehs, T. Ziehm, A. Kittel, and P. Ben-Abdallah, arXiv:1810.02628.


\bibitem{CNT1} M. Fujita, K. Wakabayashi, K. Nakada, and K. Kusakabe, J. Phys.
Soc. Jpn. {\bf 65}, 1920 (1996).

\bibitem{CNT2} K. Nakada, M. Fujita, G. Dresselhaus, and M. S. Dresselhaus, Phys. Rev. B {\bf 54}, 17 954 (1996).

\bibitem{CNT3} Ken-ichi Sasaki, Kentaro Sato, Riichiro Saito, Jie Jiang, Seiichiro Onari, and Yukio Tanaka, Phys. Rev. B {\bf 75}, 235430 (2007).

\bibitem{Recursive} C. H. Lewenkopf, and E. R. Mucciolo, J Comput Electron {\bf 12}, 203-231 (2013).

\bibitem{CNT-exp1} B. Liu, F. Wu, H. Gui, M. Zheng, and C. Zhou, ACS Nano {\bf 11}, 31 (2017).  
\bibitem{CNT-exp2} Q. Zhao, Z. Xu, Y. Hu, F. Ding, and J. Zhang, Sci. Adv. {\bf 2}, 1501729 (2016). 

\bibitem{SSH1} W. P. Su, J. R. Schrieffer, and A. J. Heeger, Phys. Rev. Lett. {\bf 42}, 1698 (1979).

\bibitem{SSH-Green} Y. Peng, Y. Bao, and F. von Oppen, Phys. Rev. B {\bf 95}, 235143 (2017).

\bibitem{non-monotonic1} A. W. Rodriguez, M.T.Homer Reid, J. Varela, J. D. Joannopoulos, F. Capasso, and S. G. Johnson, Phys. Rev. Lett. {\bf 110}, 014301 (2013).

\bibitem{non-monotonic2} H. Iizuka, and S. Fan, Phys. Rev. lett. {\bf 120}, 063901 (2018).




\end{thebibliography}
\end{document}